\documentclass[twocolumn,epsf,prl,aps]{revtex4}

\usepackage{amsfonts,amssymb,amscd,amsmath}
\usepackage{graphicx}
\usepackage{epsfig}
\usepackage{latexsym}
\usepackage{amssymb}

\def\lg#1{ | #1 \rangle }
\def\rg#1{ \langle #1 | }
\def\lr#1{ \langle #1 \rangle }
\def\lrg#1#2#3{ \langle #1 | #2 | #3 \rangle }

\newcommand{\bx}{\begin{matrix}}
\newcommand{\ex}{\end{matrix}}
\newcommand{\be}{\begin{eqnarray}}
\newcommand{\ee}{\end{eqnarray}}
\newcommand{\nn}{\nonumber \\}
\newcommand{\no}{\nonumber}
\newcommand{\de}{\delta}
\newcommand{\lt}{\left\{}
\newcommand{\rt}{\right\}}
\newcommand{\lx}{\left(}
\newcommand{\rx}{\right)}

\newcommand{\p}{\partial}

\newcommand{\al}{\alpha}

\newcommand{\vp}{\varphi}

\newcommand{\dg}{\dagger}
\newcommand{\De}{\Delta}
\newcommand{\m}{\mathrm}
\newcommand{\bn}{\bar n}

\begin{document}

\title{Super-Resolution at the Shot-Noise Limit \\ with
Coherent States and Photon-Number-Resolving Detectors}

\author{Yang Gao} \email{ygao1@tigers.lsu.edu}
\author{Christoph F.\ Wildfeuer}
\author{Petr M.\ Anisimov}
\author{Hwang Lee}
\author{Jonathan P.\ Dowling}

\affiliation{\vspace{0.1cm} \small \it Hearne Institute for
Theoretical Physics, Department of Physics and Astronomy \\
Louisiana State University, Baton Rouge, LA 70803}

\date{\today}

\begin{abstract}
There has been much recent interest in quantum optical
interferometry for applications to metrology, sub-wavelength
imaging, and remote sensing, such as in quantum laser radar (LADAR).
For quantum LADAR, atmospheric absorption rapidly degrades any
quantum state of light, so that for high-photon loss the optimal
strategy is to transmit coherent states of light, which suffer no
worse loss than the Beer law for classical optical attenuation, and
which provides sensitivity at the shot-noise limit. This approach
leaves open the question --- what is the optimal detection scheme
for such states in order to provide the best possible resolution? We
show that coherent light coupled with photon number resolving
detectors can provide a super-resolution much below the Rayleigh
diffraction limit, with sensitivity no worse than shot-noise in
terms of the detected photon power.\\

\end{abstract}

\maketitle

Ever since the work of Boto {\it et al.} in 2000, it has been
realized that by exploiting quantum states of light, such as N00N
states, it is possible to beat the Rayleigh diffraction limit in
imaging and lithography (super-resolution) while also beating the
shot-noise limit in phase estimation (super-sensitivity)
\cite{Boto2000, Kok04, Dowling2008}. 
However such quantum states of light
are very susceptible to photon loss \cite{Rubin2007, Gilbert08, Huver2008}.
Recent work has shown that in the presence of high loss, the optimal
phase sensitivity achieved is always $\De \vp = 1/\sqrt{\bn}$, where
$\bn$ is the average number of photons to arrive at the detector
\cite{Dorner2009, Demkowcz-Dobrzanski2009}. These results suggest
that, given the difficulty in making quantum states of light, as
well as their susceptibility to loss, that the most reasonable
strategy for quantum LADAR is to transmit coherent states of light,
$\lg \al=e^{-|\al|^2 /2}\sum_{k=0}^\infty (\al^k/\sqrt {k!}) \lg k$
where $\lg k$ is a $k$-photon Fock state to mitigate a super-Beer's
law in loss and maximize sensitivity \cite{Dowling2008}. It is well
known that such an approach can only ever achieve at best shot-noise
limited sensitivity \cite{Caves1981}. However, such a conclusion
leaves open the question as to what is the best detection strategy
to optimize resolution. Recent experimental results have indicated
that such a coherent-state strategy can still be super-resolving,
provided a quantum detection scheme is employed \cite{Resch2007}. In
this paper we derive an optimal quantum detection scheme, that is
super-resolving, and which can be implemented with photon number
resolving detectors. Our proposed scheme exploits coherent states of
light, is shot-noise limited in sensitivity, and can resolve
features by an arbitrary amount below the Rayleigh diffraction
limit. Our scheme would have applications to quantum optical remote
sensing, metrology, and imaging.
\begin{figure}[t!]
\begin{minipage}{0.0\textwidth}
\centerline{\epsfxsize 80mm \epsffile{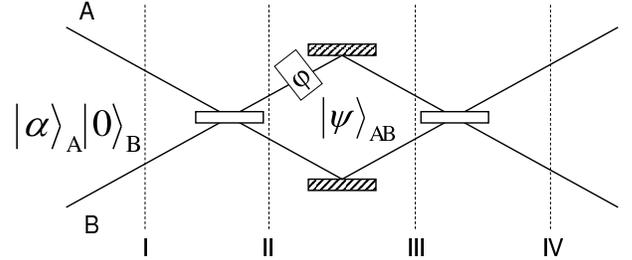}}
\end{minipage}
\caption{Here we indicate the Mach-Zehnder interferometer used in
the calculations. The coherent state is incident in mode $A$ and
vacuum in mode $B$ at the left at Line I. After the first beam
splitter transformation we have the two-mode coherent state of Eq. 
(1), as indicated at Line II. After the phase shifter $\vp$ this
state becomes the two-mode coherent state of Eq. (2). At Line III we
also implement the detection schemes corresponding to the operators
${\widehat N_{AB}}$, ${\hat \nu_{AB}}$, and ${\hat \mu_{AB}}$.
Finally after the final beam splitter, we implement the parity
operator ${\widehat \Pi_{A}}$ detection of Eq. (13) in the upper
mode.}
\end{figure}
In Fig. 1 we illustrate schematically a two-mode interferometric
quantum LADAR scheme. The source at the left is assumed to contain a
laser producing a coherent state $\lg \al_A$ in upper mode $A$ with
vacuum in lower mode $B$, which is illustrated at line I in Fig. 1.
The state is incident on a 50-50 beam splitter (BS), which mixes
this coherent state with the vacuum state $\lg 0_B$ in lower mode B.
The output of such a mixing is computed by the well known beam
splitter transformation, and is the state, \be \lg {\al/\sqrt
2,\al/\sqrt 2} = e^{-\bn/2}\sum_{n,m=0}^\infty \frac{\lx \al/\sqrt
2\rx^{n+m}}{\sqrt{n!m!}}\lg {n,m},\ee where the notation $\lg {p,q}
\equiv \lg p_A \lg q_B$, $\bn=|\al|^2$, and without loss of
generality, a BS phase factor of $i=e^{i \pi/2}$ has been dropped
for clarity. (This physically corresponds to a symmetric 50-50 BS
with a quarter-wave plate in one of the output ports.) This is the
state at line II in Fig. 1. As shown in Fig. 1 the upper mode $A$
imparts a phase shift $\vp$ on this state, yielding,
\be \lg \psi_{AB} &=& \lg {\al e^{i \vp}/\sqrt 2,\al/\sqrt 2} \\
&=& e^{-\bn/2}\sum_{n,m=0}^\infty \frac{(e^{i \vp} \sqrt{\bn /
2})^{n}(\sqrt{\bn/2})^m} {\sqrt{n!m!}} \lg {n,m}, \no \ee where,
without loss of generality, we take $\al=\sqrt{\bn}$, which amounts
to setting an irrelevant overall phase to zero. This is the state at
the line III in Fig. 1. In the coherent state basis, this state is
obviously separable. However it is less obviously separable in the
number basis, and it is easy to see that the double sum contains the
N00N states $\lg {N \!\!\!\!\!\ :: \!\!\!\!\!\ 0}^\vp=e^{i N \vp}\lg
{N, 0}+\lg {0, N}$ as well as $MM'$ states $\lg {M \!\!\!\!\  ::
\!\!\!\!\ M' }^\vp=e^{i M \vp}\lg {M,M'}+e^{i M' \vp}\lg {M', M}$,
both of which are known to be $N=M-M'$ fold super-resolving. Indeed
this observation suggests a strategy, similar to that employed by
Resch {\it et al.}, of projecting the state $\lg \psi_{AB}$ of
Eq. (2) onto the N00N basis through the implementation of the
detection operator $\widehat{N}_{AB} = \lg {N,0} \rg {0,N} + \lg
{0,N} \rg {N,0}$ \cite{Resch2007}. This results in an expectation
value of $_{AB} \lrg \psi {\widehat{N}_{AB}} {\psi} _{AB}=(\bn/2)^N
e^{-\bn} 2\cos N\vp /N!$ that is clearly $N$-fold super-resolving.
However the factor of $(\bn/2)^N e^{-\bn}/N!$ indicates that the
visibility of this expectation value, as a function of $\vp$, is
much less than unity. Let us optimize this visibility by tuning the
return power $\bn$.
The maximal visibility occurs when $\bn=N$ or when the return power
is equal to the desired super-resolution factor. For example, with
$\bn=N=10$ we achieve $10$-fold super-resolution in $_{AB} \lrg \psi
{\widehat{N}_{AB}} {\psi} _{AB}$ with a visibility of about $0.024
\%$. In a similar fashion we may now estimate the minimal phase
sensitivity variance via the usual Gaussian error propagation
formula \cite{Dowling2008}, \be \De \vp^2_N=\frac{\De
\widehat{N}_{AB}^2}{| {\p \lr {\widehat{N}_{AB}} \over \p
\vp}|^2}=\frac{2^N e^{\bn} N!-2 {\bn^N} \cos^2 N \vp}{2 \bn^N \sin^2
N\vp} \frac{1}{N^2},\ee where the factor of $1/N^2$ would provide
Heisenberg limited sensitivity, as it would be in the case of a pure
N00N state with this detection scheme \cite{Dowling2008}, if it were
not for the Poisson weight factors inherited from the coherent
state. This expression, Eq.(3), can be minimized by inspection,
again by taking $N=\bn$ and simultaneously choosing $\vp$ (which can
be done by putting a phase shifter in arm B) such that $N\vp=\pi/2$,
yielding, \be \De \vp_N^2={2^{\bn} e^{\bn} \bn! \over 2 \bn^{\bn}}{1
\over \bn^2}\cong \sqrt{{\pi \over 2}}{2^{\bn} \over
{\bn}^{3/2}},\ee where we have used Stirling's approximation, and
where again the Heisenberg limit behavior of
$1/\bn^2$ appears in tantalizing fashion. However, the exact
expression has a minimum at $\bn=2$ of $\De\vp^2_N|_{\bn=2}\cong
1.85$, and hence for all values of $\bn$ always does worse than the
shot-noise and Heisenberg limits of $\De \vp^2_\m{SNL}=1/\bn>\De
\vp^2 _\m {HL}=1/\bn^2$, in agreement with the conclusion of
\cite{Resch2007}.

It is easy to see that the reason this above strategy does worse
than the shot-noise limit (SNL), which can be achieved with simple
photon intensity difference counting \cite{Dowling2008}, is that the
projective measurement on the operator $\widehat N_{AB}$, defined
above, throws away very many photon number amplitudes in the
coherent state, keeping only the two terms in Eq. (2) with $n = N$
and $m = 0$ or $n=0$ and $m=N$. Such an observation suggests that we
introduce a new operator, which projects onto all of the maximally
super-resolving terms, that is, $\hat \nu_{AB}=\sum_{N=0}^\infty
\widehat N_{AB}$, which corresponds to the phase-bearing
anti-diagonal terms in the two-mode density matrix describing the
interferometer \cite{Huver2008}. We may now carry out the
expectation with respect to the dual-mode coherent state of Eq. (2)
to get, \be \lr {\hat \nu_{AB}} &=& 2 e^{-\bn+\bn \cos \vp
/2}\cos(\bn \sin\vp/2) \nn & \cong & 2e^{-\bn/2}\cos(\bn \vp/2),\ee
where we have approximated the expression near $\vp=0$, where the
function is sharply peaked. We can see from this expression that by
carefully choosing the return power $\bn =N$, we recover super
resolution. However the exponential pre-factor produces a loss in
visibility. Choosing $\bn=N=20$, to obtain again $10$-fold
super-resolution, we obtain a visibility of only $0.009\%$. This low
visibility does not bode well for the sensitivity, which we may
compute as per Eq. (3) to get, \be \De \vp_\nu^2 &=& \lt
e^{\bn}+e^{3\bn/2} -e^{\bn \cos\vp}[1+\cos(\bn \sin\vp)] \rt \nn &&
\times e^{-\bn \cos \vp} \csc^2\lx \vp+{\bn \over 2} \sin \vp \rx {2
\over \bn^2},\ee which again displays the tempting
Heisenberg-limiting pre-factor of $1/\bn^2$. This expression is
singular at the phase origin, but has a minimum nearby
$\vp=\pi/2\bn$, which for large photon number is approximately at
$\bn=N$, which by again choosing this to be a large integer,
simplifies Eq. (6) to, \be \De \vp_\nu^2|_{\vp=\pi/2\bn} \cong {16\pi
e^{\bn/2}\over (2\bn+\pi)^2},\ee which only approaches the SNL and
HL near $\bn=1$ and then rapidly diverges to be much worse than the
SNL for large $\bn$. Hence, from these examples, we see there is a
high price to pay for N00N-like super-resolution with coherent
states --- so many photon amplitudes are discarded that we always do
far worse than shot noise. This analysis then suggests our final
protocol --- what if we choose a measurement scheme that includes
all of the phase-carrying off-diagonal terms in the two-mode density
matrix? Such a scheme is to consider the operator constructed from
all the $MM'$ projectors \cite{Huver2008}, that is, \be
\hat{\mu}_{AB}=\sum_{M,M'=0}^\infty \lg {M',M}\rg {M ,M'} ,\ee where
we note this is evidently not a resolution of the identity operator.
It is easy to show that this operator of Eq. (8) is both Hermitian
and idempotent, that is $\hat{\mu}_{AB}^\dg =\hat{\mu}_{AB}$ and
$\hat{\mu}_{AB}^2=\hat{I}_{AB}$, respectively, where $\hat{I}_{AB}$
is the two-mode identity operator. Using these properties, with a
bit of algebra, we establish, with respect to the two-mode coherent
state of Eq. (2), the following results, \be \lr {\hat{\mu}_{AB}} &=&
e^{-2\bn \sin^2 (\vp/2)} ,\\ \De \vp^2_\mu &=&
{e^{4\bn\sin^2(\vp/2)}-1 \over \bn^2 \sin^2 \vp}, \ee where once
again the Heisenberg limit term of $1/\bn^2$ appears accompanied by
an exponential factor in the sensitivity estimate. We plot as a
solid curve the expectation value of Eq. (9), for a return power of
$\bn=100$, in Fig. 2, along with the standard (classical) photon
difference detection interferogram (dashed curve)
\cite{Dowling2008}. Clearly $\lr{\hat{\mu}_{AB}}$ has a visibility
of $100 \%$ now, and is periodic in $\vp$ with period $2\pi$, and
highly peaked at the phase origin where $\vp=0$. This curve is not
super-resolving in the usual sense of the word, as there are no
multiple narrow peaks as would be the case in a N00N-state scheme,
but it is super-resolving in the sense that there is a well defined
narrow feature that is clearly sub-Rayleigh limited in resolution.
Such a feature would be useful, for example, in LADAR ranging or
laser Doppler velocimetry, where one would lock onto the side of
such a feature and then monitor how it changes in time with a
feedback loop in the interferometer.
\begin{figure}[t!]
\begin{minipage}{0.0\textwidth}
\centerline{\epsfxsize 70mm \epsffile{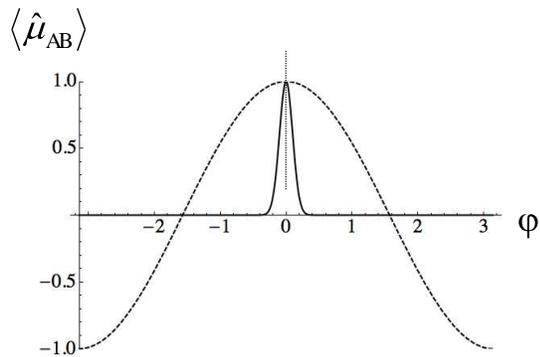}}
\end{minipage}
\caption{This plot shows the expectation value $\lr {\hat \mu_{AB}}$
of Eq. (9) plotted as a function of the phase shift $\vp$ (solid
curve) for a return power of $\bn=100$. For reference we plot the
normalized ``classical" two-port difference signal (dashed curve).
We see that the plot of the $\lr {\hat \mu_{AB}}$ is super-resolving
and is narrower than the classical curve by a factor of $\de \vp
=1/\sqrt{\bn} =1/10$, as given in Eq. (11).}
\end{figure}

\begin{figure}[t!]
\begin{minipage}{0.0\textwidth}
\centerline{\epsfxsize 70mm \epsffile{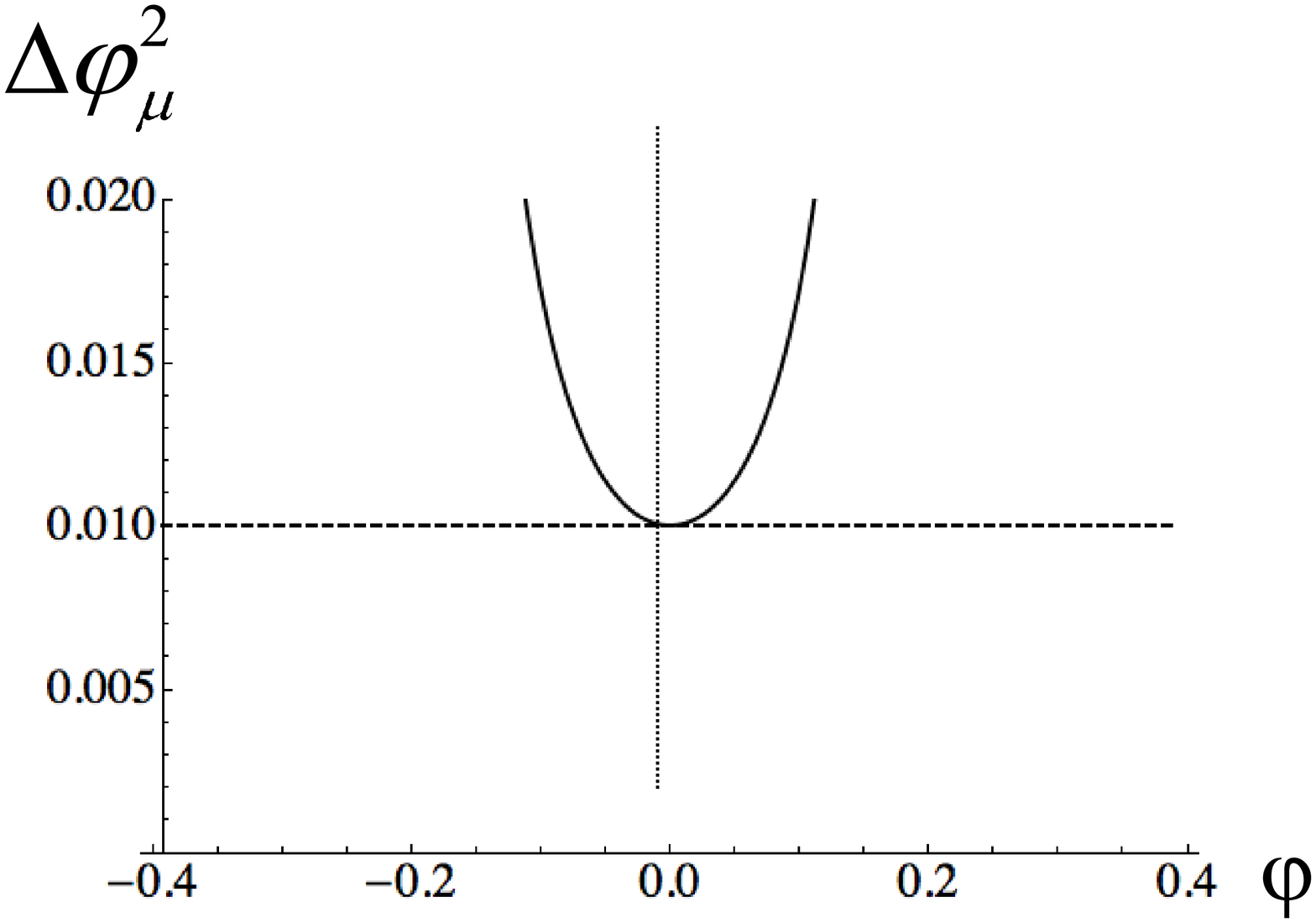}}
\end{minipage}
\caption{In this plot we depict the sensitivity expression $\De
\vp_\mu^2$ of Eq. (10), again for the return power of $\bn=100$
(solid curve). The horizontal dashed line indicates the shot-noise
limit of $\De \vp_\m{SNL}^2=1/\bn =1/100$. We see that the
sensitivity of the super-resolving $\hat \mu_{AB}$ detection scheme
hits the SNL at $\vp=0$, as indicated by expanding Eq. (10) in a
power series. }
\end{figure}
To estimate the width of this central peak we note that in the small
phase angle limit, Eq. (9) may be approximated as, \be \lr
{\hat{\mu}_{AB}}|_{\vp \cong 0}\cong e^{-\bn \vp^2/2} ,\ee which is
clearly a Gaussian of width $\de \vp = 1/ \sqrt {\bn}$. Hence by
choosing a return power of $\bn=100$, we are $10$-fold super
resolving in terms of this central feature of the expectation value.

Now we check the sensitivity of this scheme. In Fig. 3 we plot as a
solid curve the variance of Eq. (10), again for $\bn=100$, near the
phase origin over $-\pi/8<\vp<\pi/8$. We include as a dashed curve
the shot-noise limit. We see that the sensitivity is shot-noise
limited about the phase origin. This may be established analytically
by first noting that the expression of Eq. (10) for the sensitivity
has a removable singularity at the origin, and then by expanding it
in a power series around $\vp=0$ to get $\de \vp_\mu|_{\vp \cong 0}
\cong 1/\sqrt {\bn} $, which is precisely the shot-noise limit.
Hence by counting all the photons in an off-diagonal fashion, the
detection scheme embodied in the operator $\hat{\mu}_{AB}$ of Eq. 
(8) produces a new kind of super resolution, and performs at the
shot-noise limit in sensitivity. Since all of this information is
extracted at the detector, and only coherent states are used at the
source and in the interferometer, this scheme will be no worse in
sensitivity in the presence of absorption or loss than an equivalent
classical LADAR scheme, but it will in addition have super-resolving
capabilities.

It remains to be understood how the observable of Eq. (8) might be
detected in the laboratory. We note that at this stage of the
analysis the operator $\hat{\mu}_{AB}$ is to be carried out with
respect to the two-mode coherent state at the line III in the Fig. 1.
That is we have not yet applied the second beam splitter.
To the right of the second beam splitter, at the line IV in Fig. 1,
we wish to now carry out the well-known parity operator \cite{Gerry00}
on output mode $A$ in the upper arm, 
\be 
\widehat{\Pi}_A=(-1)^{\hat n_A}=e^{i \pi \hat a^\dg \hat a} ,
\ee which simply indicates whether an even
or odd number of photons exits that port. 
Here $\hat n_A=\hat a^\dg
\hat a$ is the number operator for that mode. Such a detection
scheme is easily implemented by placing a highly efficient
photon-number-resolving detector at this port, and such detectors
with $95\%$ efficiency and number resolving capabilities in the tens
of photons have been demonstrated \cite{Wildfeuer2009a}. The
connection between these two operators may be established via the
easily proved identity, \be _{AB} \lrg \psi {\hat{\mu}_{AB}} \psi
_{AB} \equiv \lrg {\al,0} {U^\dg (\widehat{\Pi}_{A}\otimes \hat I_B)
U} {\al,0}, \ee where $\hat I_B$ is the $B$-mode identity operator,
and $U=e^{-i \hat{J}_y \vp}$ with
$\hat{J}_y={i}(\hat{a}\hat{b}^\dg-\hat{a}^\dg \hat{b})/2$ denotes
the transformation of the usual Mach-Zehnder interferometer. That
is, the effect of measuring the two-mode coherent state with respect
to the operator $\hat{\mu}_{AB}$ to the left of the second BS at
line III is equivalent to measuring it with respect to the parity
operator $\widehat{\Pi}_{A}$ to the right of the second BS at line
IV, all indicated in Fig. 1. Hence Eq. (14) establishes that the two
schemes have the same super-resolving and shot-noise limiting
properties, with the important point that the parity operator is
perhaps far easier to implement in the laboratory, and it has been
show to be a universal detection scheme in quantum interferometry
\cite{Campos03, Gao2008}.

In summary we have provided a super-resolving interferometric
metrology strategy, which achieves the shot-noise limit. The
protocol has the appealing feature that it requires only the
production and transmission of ordinary laser beams in the form of
coherent states of light. Hence, unlike the issues concerning the
propagation of non-classical states of light, such as squeezed light
or entangled Fock states, this scheme clearly suffers no worse
degradation in the presence of absorption and loss than a classical
coherent LADAR system. All of the quantum trickery, which provides
the super resolution, is carried out in the detection, which can be
carried out with current photon number-resolving technology.
However, a scheme by which we count the number of photons and then
decide if that number is even or odd is overkill. The parity
operation only requires that we know the sign --- even or odd ---
independently of the actual number. Hence counting photon number is
sufficient but perhaps not necessary, if a general scheme to
determine the parity of a photon state could be found that did not
require photon number counting. We conjecture such a scheme exists,
perhaps through the exploitation of optical nonlinearities \cite{Gerry02},
or projective measurements, and this is an area of ongoing research.
Our protocol for super-resolving phase measurements at the
shot-noise limit has applications to quantum imaging, metrology, and
remote sensing. In particular, for applications such as Doppler
velocimetry of rapidly moving objects, one does not have the luxury
to integrate the data for long periods of time in order to push down
the signal to noise. In such scenarios the signal resolution in the
form of the Rayleigh criteria is the usual limit to the system
performance. It is in such situation that we anticipate this
protocol to be most useful.

We would like to acknowledge interesting and useful discussions with
Alan Migdall, Seth Lloyd, Sae-Woo Nam, and Jeffrey Shapiro. We would
also like to acknowledge support from the Army Research Office, the
Boeing Corporation, the Defense Advanced Research Projects Agency,
the Foundational Questions Institute, the Intelligence Advanced
Research Projects Activity, and the Northrop-Grumman Corporation.

\end{document}